\documentclass{article}

\usepackage{microtype}
\usepackage{graphicx}
\usepackage{subfigure}
\usepackage{booktabs} 
\usepackage{multirow}
\usepackage{makecell}
\usepackage{threeparttable}
\usepackage{physics}

\usepackage{hyperref}

\usepackage[accepted]{icml2025}

\usepackage{amsmath}
\usepackage{amssymb}
\usepackage{mathtools}
\usepackage{amsthm}

\usepackage[capitalize,noabbrev]{cleveref}

\theoremstyle{plain}

\theoremstyle{definition}

\theoremstyle{remark}

\usepackage[textsize=tiny]{todonotes}

\icmltitlerunning{eccDNAMamba: A Pre-Trained Model for Ultra-Long eccDNA Sequence Analysis}

\begin{document}

\twocolumn[
\icmltitle{eccDNAMamba: A Pre-Trained Model for Ultra-Long eccDNA Sequence Analysis}



\icmlsetsymbol{equal}{*}

\begin{icmlauthorlist}
\icmlauthor{Jien Li}{equal,bio}
\icmlauthor{Zhenke Liu}{equal,cs}
\icmlauthor{Ziqi Zhang}{equal,cs}
\end{icmlauthorlist}

\icmlaffiliation{cs}{Department of Computer Science, Brown University, Providence, RI, USA}
\icmlaffiliation{bio}{Department of Molecular Biology, Cell Biology, and Biochemistry, Brown University, Providence, RI, USA}
\icmlcorrespondingauthor{Zhenke Liu}{zliu328@brown.edu}

\icmlkeywords{Machine Learning, ICML}

\vskip 0.3in
]



\printAffiliationsAndNotice{\icmlEqualContribution} 

\begin{abstract}
Extrachromosomal circular DNA (eccDNA) plays key regulatory roles and contributes to oncogene overexpression in cancer through high-copy amplification and long-range interactions. Despite advances in modeling, no pre-trained models currently support full-length circular eccDNA for downstream analysis. Existing genomic models are either limited to single-nucleotide resolution or hindered by the inefficiency of the quadratic attention mechanism. Here, we introduce eccDNAMamba, the first bidirectional state-space encoder tailored for circular DNA sequences. It combines forward and reverse passes for full-context representation learning with linear-time complexity, and preserves circular structure through a novel augmentation strategy. Tested on two real-world datasets, eccDNAMamba achieves strong classification performance and scales to sequences up to 200 Kbp, offering a robust and efficient framework for modeling circular genomes. Our codes are available at \url{https://github.com/zzq1zh/GenAI-Lab.git}.

\end{abstract}

\section{Introduction}

Extrachromosomal circular DNA (eccDNA) has emerged as a ubiquitous and functionally important element in diverse organisms and conditions, with a particularly pronounced relevance in cancer genomes.
These circular DNA molecules, which range from a few hundred base pairs to over megabase, often carry oncogenes or distal regulatory elements and contribute to tumor evolution, therapeutic resistance, and intratumor heterogeneity~\cite{ecc1,ecc2,ecc3}. 

In biomedical applications, eccDNA has been implicated in diverse processes such as oncogene amplification, genome instability, and non-Mendelian inheritance of drug resistance. 
Recent studies suggest that understanding the structure and function of eccDNA can provide novel insights into cancer diagnostics and personalized treatment strategies~\cite{ecclin1,ecclin2,ecclin3}.

Despite advances in eccDNA detection pipelines (e.g., eccDNA-Pipe ~\cite{10.1093/bib/bbae034}), analyzing the full-length sequence of eccDNA remains a significant challenge. 
The circular nature of eccDNA introduces unique modeling constraints: linearizing sequences at arbitrary breakpoints creates artificial boundaries, potentially disrupting biologically meaningful head–tail interactions~\cite{ling2021eccdna}. Additionally, many eccDNAs exceed 10,000 bp in length, which precludes standard Transformer architectures due to their quadratic complexity~\cite{zhou2023dnabert}. 
These structural and computational challenges hinder the development of principled models that reason over the full circular topology.

Recent genomic foundation models demonstrate both promise and limitation. HyenaDNA~\cite{hyenaDNA} leverages implicit convolutions for long sequences but remains unidirectional and topology-agnostic. 
DNABERT-2~\cite{zhou2023dnabert} uses byte-pair encoding to capture motifs but retains a linear input assumption and standard Transformer layers.
Moreover, state-space models such as
Caduceus offers linear scalability~\cite{schiff2024caduceus} but focuses on single-nucleotide resolution and restrain their scalability on ultra-long eccDNA sequences for downstream applications. 
These limitations prevent modeling ultralong
eccDNA sequences.

To address these challenges, we propose eccDNAMamba, the first bidirectional state-space encoder tailored for circular DNA. 
Our design fuses forward and reverse passes of a Mamba-2 backbone to provide full contextual encoding while preserving linear time complexity.
We introduce an augmentation strategy that preserves head–tail dependencies by appending 64 tokens from the start of each sequence to its end. Training is performed using a SpanBERT-style objective~\cite{joshi2020spanbert}, which masks contiguous spans to encourage motif-level reconstruction.

Our proposed model, eccDNAMamba, demonstrates strong and stable performance across multiple classification tasks, including cancer-state prediction and distinguishing authentic eccDNAs from synthetic circular fragments, consistently outperforming domain-specific baselines.
Notably, it generalizes to full-length sequences up to 200,000 base pairs without performance degradation, underscoring its scalability for ultra-long circular genome modeling.
These results position eccDNAMamba as a robust foundation for downstream applications in eccDNA analysis.

\section{Related Work}

\subsection{Genomic Foundation Models}
The past few years have seen the rise of genomic foundation models that adapt natural language processing pretraining paradigms to DNA. 
DNABERT-2~\cite{zhou2023dnabert} introduced byte-pair encoding (BPE) tokenization to capture sequence motifs, while the Nucleotide Transformer ~\cite{mendoza2024foundational} scaled pretraining to multiple genomes. 

There also exist other genomic foundation models that can model ultra-long sequences. HyenaDNA~\cite{hyenaDNA} leverages implicit convolutions for long sequences. Caduceus offer linear scalability~\cite{schiff2024caduceus}.

eccDNAMamba differs from this paradigm by combining the scalability of state-space models with a bidirectional encoder architecture and span masking training objective. 
This allows us to capture global dependencies while maintaining compatibility with span-masked objectives and motif-centric tokenization. 

\subsection{eccDNA Modeling}
Most computational efforts on eccDNA to date have focused on detection, breakpoint localization, and cataloging. 
Tools such as eccDNA-Pipe~\cite{10.1093/bib/bbae034} identify circular structures through split-read and discordant alignment signatures but do not attempt sequence-to-function modeling. 
Recent work has explored the use of deep learning for eccDNA classification. 
For instance, Li et al. ~\cite{10.1093/bib/bbad147} proposed a CNN-based classifier that identifies eccDNA from short non-eccDNA sequence segments.

In contrast, eccDNAMamba performs self-supervised pretraining over long circular sequences, capturing long-range dependencies and structural context across diverse eccDNA types. 

\subsection{Mamba and State-Space Models}

Mamba~\cite{gu2023mamba} is a recently proposed state-space model (SSM) that models input sequences through a discretized linear dynamical system as shown in Figure ~\ref{fig: Model Architecture}:

\begin{equation}
h_t = A h_{t-1} + B x_t, \quad y_t = C h_t
\end{equation}

where \( x_t \in \mathbb{R}^d \) is the input at position \( t \), \( h_t \in \mathbb{R}^d \) is the hidden state, \( y_t \in \mathbb{R}^m \) is the output, and \( A, B, C \) are learnable matrices.  

To improve expressiveness, Mamba introduces input-dependent gating and parameterizes output as:

\begin{equation}
y_t = \sum_{j=1}^{t} K_{t-j} \odot (G_j \odot x_j)
\end{equation}

where \( K \) is a learned convolutional kernel derived from the state-space dynamics, \( G_j \) is a gating function, and \( \odot \) denotes element-wise multiplication.  
This structure allows long-range dependency modeling while maintaining linear time and memory complexity.  

Mamba-2~\cite{dao2024transformers} builds upon this by introducing Dynamic Block Sparse Selective Scans as illustrated in Figure ~\ref{fig: Model Architecture}.
It reformulates the SSM kernel using semi-separable matrices and applies block-wise sparsity, significantly improving runtime efficiency.  

We adopt Mamba-2 as the base encoder in our framework and extend it to support bidirectional processing.  
Unlike the original decoder-only formulation, our adaptation enables full-context encoding over circular DNA while preserving the model's linear scalability.

\section{Method}

\subsection{The eccDNAMamba Model}
\label{sec: model}
\begin{figure*}[htb]
    \centering
    \includegraphics[width=0.97\textwidth]{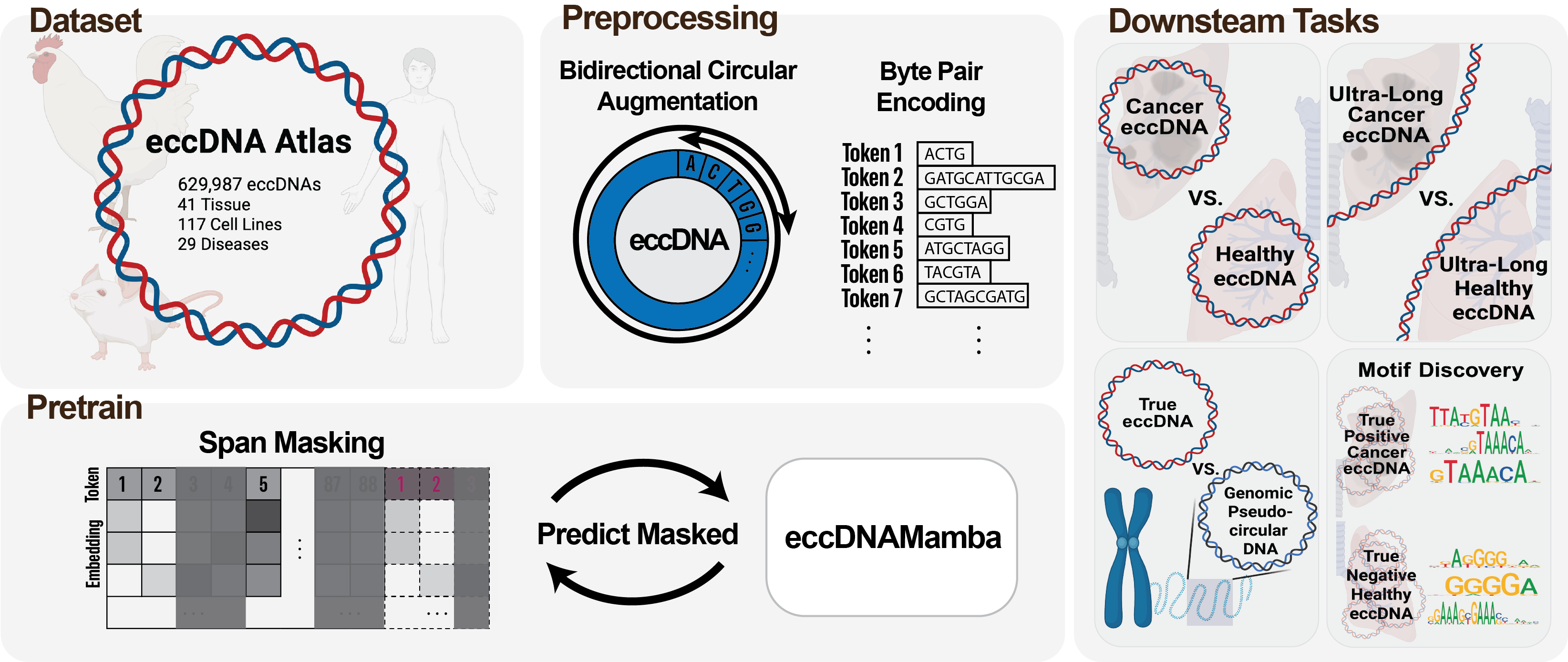}
    \caption{eccDNAMamba initially embeds the input eccDNA sequence and incorporates circular augmentation to fit their periodic structure. The embedded representations are subsequently processed by Mamba blocks to capture and learn intricate patterns within the eccDNA sequences.}
    \label{fig:Model Overview}
\end{figure*}

As shown in Figure~\ref{fig:Model Overview}, given a BPE-tokenized eccDNA input sequence
\begin{align}
x = (x_1, x_2, \dots, x_L), \quad x_i \in \mathcal{V}
\end{align}

where $\mathcal{V}$ denotes the vocabulary constructed via Byte-Pair Encoding (BPE). The model appends a special \texttt{[CLS]} token to the front of the input sequence to provide a unified semantic representation of the entire input sequence. To perform circular augmentation, the first $s = 64$ tokens are appended to the end of the sequence, resulting in:
\begin{align}
\tilde{x} = (\texttt{[CLS]}, x_1, x_2, \dots, x_L, x_1, x_2, \dots, x_s)
\end{align}

\begin{figure*}[htb]
    \centering
    \includegraphics[width=0.6\textwidth]{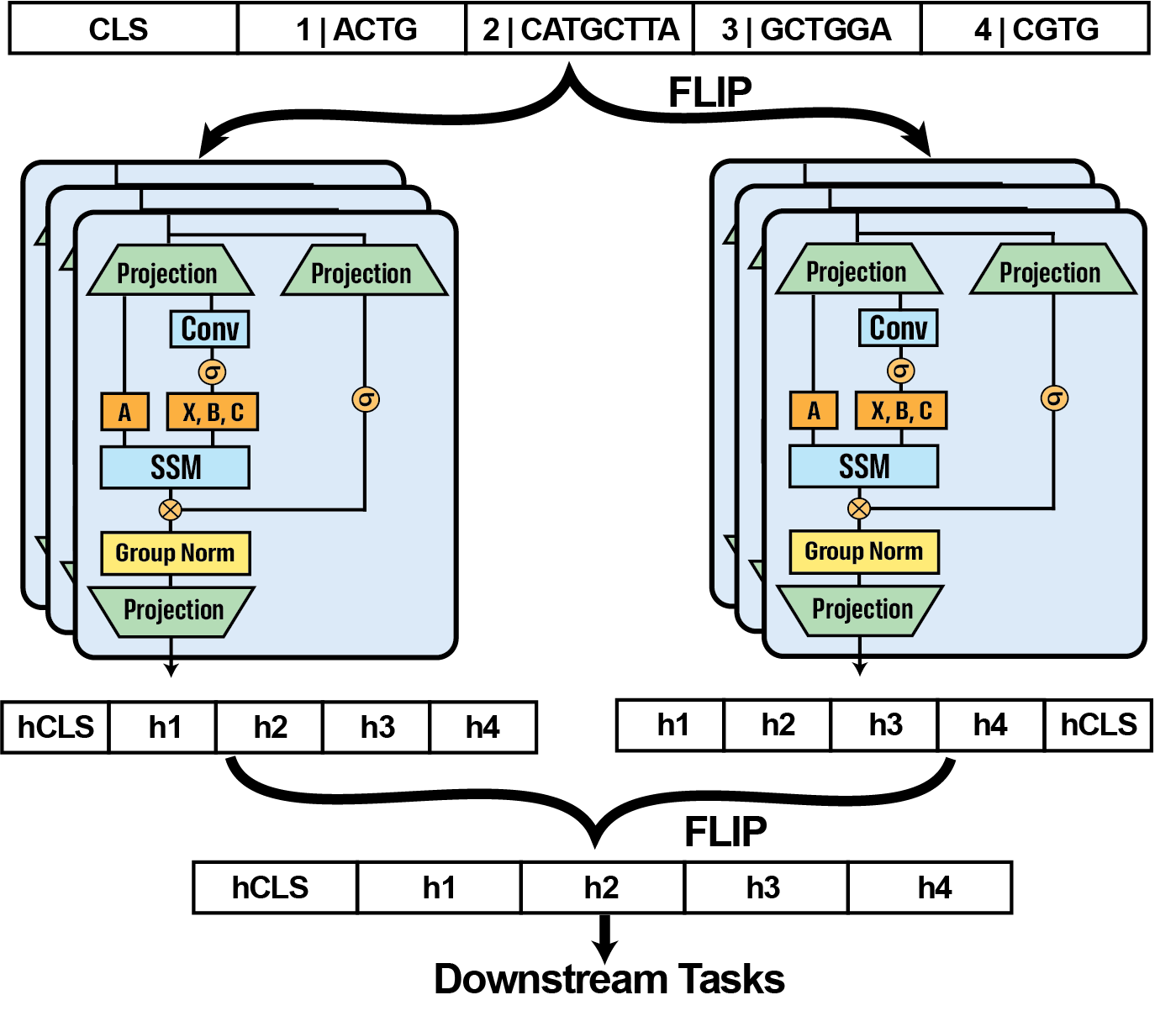}
    \caption{eccDNAMamba Architecture.}
    \label{fig: Model Architecture}
\end{figure*}

The embedded sequence is then fed into two directional Mamba-2 encoders to model bidirectional contextual information. Here, $\mathbf{h}$ denotes the contextual embedding. We use $\overset{\rightarrow}{(\cdot)}$ and $\overset{\leftarrow}{(\cdot)}$ to denote the forward and backward directions, respectively:
\begin{align}
\overset{\rightarrow}{\mathbf{h}} &= \overset{\rightarrow}{\text{Mamba}}(\overset{\rightarrow}{\mathbf{\tilde{x}}}) = [\overset{\rightarrow}{h}_{cls}, \overset{\rightarrow}{h}_1, \overset{\rightarrow}{h}_2, \ldots, \overset{\rightarrow}{h}_L] \\
\overset{\leftarrow}{\mathbf{h}} &= \overset{\leftarrow}{\text{Mamba}}(\overset{\leftarrow}{\mathbf{\tilde{x}}}) = [\overset{\leftarrow}{h}_L, \overset{\leftarrow}{h}_{L-1}, \ldots, \overset{\leftarrow}{h}_1, \overset{\leftarrow}{h}_{cls}]
\end{align}


where $\vec{h}_{\text{cls}}$ denotes the output corresponding to the special \texttt{[CLS]} token for global sequence representation, and $\vec{h}_i$ represents the contextual embedding of the $i$-th input token ($1 \leq i \leq L$), with $L$ being the input sequence length. The model aligns and concatenates the two directional outputs, and integrates them through a shared MLP layer to produce the final output embeddings, as illustrated in Figure~\ref{fig: Model Architecture}. 

This bidirectional structure allows the model to learn both upstream and downstream circular dependencies. The resulting output embeddings can be used in downstream tasks. In this work, we use the contextual embedding corresponding to the \texttt{[CLS]} token from the final output as the input for downstream tasks. A two-layer MLP is then used for classification training and prediction.

\subsection{Components of eccDNAMamba}
\subsubsection{Span Masking}
In this study, we adopt the Masked Language Modeling (MLM) task used in BERT for pre-training and further enhance the pretraining performance of eccDNAMamba by incorporating improvements proposed in SpanBERT~\cite{joshi2020spanbert}.

The traditional training approach for the BERT Masked Language Model (MLM) task involves randomly selecting 15\% of the tokens in each sample. Among these selected tokens, 80\% are replaced with the special \texttt{[MASK]} token, 10\% are replaced with a random token, and the remaining 10\% are left unchanged. In the label vector, the known (unmasked) tokens are masked out so they do not contribute to the loss computation. The model then focuses solely on the difference between the masked tokens and their corresponding ground truth values. Finally, the model uses the crossEntropyLoss to calculate the discrepancy between the predictions and the actual tokens.

As shown in Figure~\ref{fig:Model Overview}, we adopt the improved SpanBERT strategy, which replaces the random selection of individual tokens with the selection of random contiguous spans, with each span averaging around 3 consecutive tokens. The total number of selected tokens accounts for 15\% of the sequence. These spans are then randomly replaced using: 80\% \texttt{[MASK]} tokens, 10 \%random tokens, and 10\% unchanged tokens. This training strategy requires the model to predict masked spans rather than isolated tokens. 

\subsubsection{Byte-Pair Encoding}

DNA sequences are unstructured and must be tokenized before input to deep learning models.

We adopt Byte Pair Encoding (BPE) ~\cite{sennrich2015neural, gage1994new}, which adaptively merges frequent adjacent substrings:
\begin{align}
(a^*, b^*) &= \underset{(a, b)}{\arg\max} \; \text{freq}_{C_t}(a, b) \\
\mathcal{V}_{t+1} &= \mathcal{V}_t \cup \{a^*b^*\}
\end{align}
Here, $C_t$ denotes the current corpus and $\mathcal{V}_t$ the vocabulary at iteration $t$. The BPE algorithm iteratively merges the most frequent adjacent pair of symbols $(a^*, b^*)$, where the superscript $^*$ indicates that the pair is selected as the most frequent in the current iteration. The merged token $a^*b^*$ is then added to the vocabulary. Through this process, BPE identifies high-frequency nucleotide patterns and encodes them as tokens, allowing the model to operate directly on motif-like structures rather than individual nucleotides. This significantly reduces the total number of tokens in a given sequence, enabling the model to handle longer input sequences more efficiently. We provide evidence in Appendix Section~\ref{sec:data-2} to show the compression rate of BPE encoding.


\subsubsection{Circular Data Augmentation}

eccDNAs are circular molecules, yet standard linear representations ignore potential head–tail dependencies. Prior studies suggest the ends of eccDNA molecules often contain repetitive elements involved in circularization ~\cite{zhao2022extrachromosomal}. To preserve this structure, we introduce a circular augmentation strategy: appending the first 64 tokens of a sequence to its end and we provide rationale in section ~\ref{sec:data-1}. For longer sequences, it enables learning of long-range dependencies between informative prefix regions and the tail. We empirically validate the benefits of this augmentation in Appendix Section~\ref{sec:ablation-1}.

\subsubsection{Bidirectionality}

The Mamba-2 architecture was originally designed for GPT-style tasks and follows a decoder-only structure, which allows information flow in only one direction from past tokens. However, our goal is to train an encoder-only model using a masked language modeling (MLM) task, which requires bidirectional context awareness.

To enable this, we modified the Mamba-2 architecture to incorporate bidirectional perception as shown in Figure~\ref{fig: Model Architecture} ~\cite{huang2015bidirectional}. Specifically, we simultaneously train two instances of the Mamba-2 model: one processes the forward sequence, while the other processes the reversed sequence. During each forward pass, the hidden state outputs from both directions are aligned to the forward order and aggregated using a shared MLP layer.

Through this aggregation, the model can integrate information from both directions and learn a global representation of the input sequence.
We enable Mamba-2 to be trained on the MLM task and function as an encoder for generating latent token representations.

\subsubsection{Padding}

Since Mamba-2 is designed for causal masking, it lacks native support for padding. Consequently, introducing padding during training may inject noise into hidden states. To mitigate this, we adopt two strategies: (1) we zero out the embeddings of padding tokens and prevent their updates during training, and (2) we apply Transformer-style attention masks to suppress the influence of padding. Additionally, we reset hidden states and residuals at padding positions to zero in every layer.

\section{Experiments}

In section ~\ref{sec:data-pre} and ~\ref{sec:pre}, we introduce our pretraining on the eccDNA sequences. We introduce baseline models and evaluation metric in ~\ref{sec:baseline} and ~\ref{sec:metircs}. In section ~\ref{sec:task1} and ~\ref{sec:task2}, we evaluate eccDNAMamaba's performance on classification tasks against state of art(SOTA) models. 

\subsection{Data}
\label{sec:data-pre}

As shown in Table~\ref{tab:eccDNA}, we pre-train our model on a curated corpus of 120{,}000 eccDNA sequences, comprising approximately 100 million tokens. These sequences are sampled from two large-scale resources: CircleBase~\cite{zhao2022circlebase} and eccDNA Atlas~\cite{zhong2023eccdna}. 
CircleBase focuses exclusively on human eccDNA and contains 601,036 annotated sequences, while eccDNA Atlas aggregates $629,987$ eccDNAs from multiple animal species. 
We leverage both datasets to pre-train and evaluate eccDNAMamba, aiming to capture both human-specific and cross-species eccDNA patterns. 

To accelerate pre-training, we filter out sequences longer than 10kbp, resulting in a final corpus of $1,087,886$ eccDNA sequences totaling $524$ million base pairs. 
After applying BPE tokenization, this corpus comprises approximately $101.5$ million tokens, yielding an average token length of $5.16$ base pairs per token after BPE tokenization. 

\begin{table*}[!htbp]
\tabcolsep=0pt
\begin{tabular*}{\textwidth}{@{\extracolsep{\fill}}lccc@{\extracolsep{\fill}}}
\toprule%
Dataset & Species & Number of eccDNAs & Additional Information \\
\midrule
CircleBase~\cite{zhao2022circlebase} & Human & 601,036 & Includes genes, regulatory elements, etc. \\
eccDNA Atlas~\cite{zhong2023eccdna} & Animals & 629,987 & Disease, expression, and regulatory annotations \\
\bottomrule
\end{tabular*}
\begin{tablenotes}%
\item\caption{List of eccDNA datasets used in this study.\label{tab:eccDNA}}
\end{tablenotes}
\end{table*}
\subsection{Pretraining}
\label{sec:pre}

We adopt a hybrid method combining contextual embedding-level control and masking as reported in previous section during pre-training. 

We pre-train the model using a span-masked language modeling (SpanMLM) objective adapted for circular DNA modeling. 
Specifically, we use a bidirectional variant of the Mamba architecture, denoted as \textbf{BiMambaForMaskedLM}, initialized from the \texttt{state-spaces/mamba2-130m} configuration. 
The model is trained from scratch on a custom eccDNA corpus, tokenized using a byte-pair encoding vocabulary. The tokenizer includes the special tokens \texttt{[PAD]}, \texttt{[UNK]}, \texttt{[MASK]}, and \texttt{[CLS]}.

Let the input sequence be denoted as $$x = (x_1, x_2, \dots, x_L)$$ and let $\mathcal{M} \subset \{1, \dots, L\}$ be a set of span-masked positions. 
The model is trained to reconstruct the original tokens at these positions by minimizing the span-masked cross-entropy loss:
$$
\mathcal{L}_{\text{SpanMLM}} = - \sum_{i \in \mathcal{M}} \log p_\theta(x_i \mid x_{\backslash \mathcal{M}})
$$
where $p_\theta$ denotes the model’s conditional output distribution and $x_{\backslash \mathcal{M}}$ denotes the input with masked spans.

Training is conducted using the HuggingFace \texttt{Trainer} API. We set the effective batch size to 144 by using 6 samples per GPU and accumulating gradients over 8 steps. 
The model is trained for 3 epochs using a learning rate of $5 \times 10^{-4}$, with a linear warmup over 6\% of total steps. We use the AdamW optimizer with a weight decay of 0.01. 
All training runs use BF16 mixed precision and are executed on a single node with 3 NVIDIA L40S GPUs.

\subsection{Baseline}
\label{sec:baseline}

 For the task of classifying eccDNA of diseased and healthy origins in Section  ~\ref{sec:task1}, to benchmark performance, we compared eccDNAMamba against the state-of-the-art model DNABERT-2, HyenaDNA, and Caduceus using officially recommended learning rates. For task differentiating eccDNA from genomic pseudo-circular DNA
 ~\ref{sec:task2}, we applied the same setting for eccDNAMamba and adopt the officially recommended hyperparameters for DeepCircle.~\cite{10.1093/bib/bbad147}

\subsection{Metrics}
\label{sec:metircs}
We employed the standard cross-entropy loss to train all classification tasks. Given an input sequence $x$ and ground-truth label $y \in {1, \dots, C}$, the model produces a predicted probability distribution $\hat{p} = \text{softmax}(f_\theta(x))$, where $f_\theta$ is the task-specific classification head. The loss is computed as:
$$
\mathcal{L}_{\text{CE}} = - \sum_{c=1}^{C} \mathbb{1}(y = c) \cdot \log \hat{p}_c
$$
We report four standard classification metrics: precision, recall, accuracy, and macro F1. These metrics are computed based on the number of true positives ($TP$), true negatives ($TN$), false positives ($FP$), and false negatives ($FN$), where $TP$ and $TN$ refer to correctly predicted positive and negative samples, respectively, while $FP$ and $FN$ correspond to incorrect predictions.

\textbf{Precision} measures the proportion of predicted positives that are correct:
$$
\text{Precision} = \frac{TP}{TP + FP}
$$
\textbf{Recall} measures the proportion of actual positives that are correctly predicted:
$$
\text{Recall} = \frac{TP}{TP + FN}
$$
\textbf{Accuracy} measures the overall proportion of correct predictions:
$$
\text{Accuracy} = \frac{TP + TN}{TP + TN + FP + FN}
$$
\textbf{Macro-F1} is the unweighted average of per-class F1 scores:
$$
\text{Macro-F1} = \frac{1}{C} \sum_{i=1}^{C} \text{F1}_i
$$
\begin{table*}[!htbp]
\centering
\renewcommand{\arraystretch}{1.2}
\resizebox{\textwidth}{!}{
\begin{tabular}{llllcccc}
\hline
\textbf{Task} & \textbf{Model} & \textbf{Training set (seq)} & \textbf{Test set (seq)} & \textbf{F1} & \textbf{accuracy} & \textbf{precision} &\textbf{recall} \\
\hline
\multirow{2}{*}{Cancer vs.~Healthy(\textless10\,kb)} 
& \textbf{eccDNAMamba} & 20,000 (10,000 cancer + 10,000 healthy) & 4,000 & \textbf{0.8242} & \textbf{0.8242} & 0.8242 & \textbf{0.8242} \\
& DNABERT-2            & 20,000 (10,000 cancer + 10,000 healthy) & 4,000 & 0.8187        & 0.8187 & 0.8187 & 0.8187 \\
& HyenaDNA            & 20,000 (10,000 cancer + 10,000 healthy) & 4,000 & 0.8105	& 0.8104	& 0.8105	& 0.8105 \\
& Caduceus  & 20,000 (10,000 cancer + 10,000 healthy) & 4,000 & 0.8216	& 0.822 &	\textbf{0.8248}	& 0.822 \\
\hline
\multirow{2}{*}{Cancer vs.~Healthy (10–200 kb)}
& \textbf{eccDNAMamba} & 2,000 (1,000 cancer + 1,000 healthy)   & 400   & \textbf{0.8147} & \textbf{0.8175} & \textbf{0.8377} & \textbf{0.8174} \\
& DNABERT-2            & 2,000 (1,000 cancer + 1,000 healthy)  & 400 & 0.5702 &	0.5725 	& 0.574 & 	0.5725 \\
& HyenaDNA            & 2,000 (1,000 cancer + 1,000 healthy)  & 400 & 0.7261 &	0.735	& 0.7699	& 0.735 
\\
& Caduceus  & 2,000 (1,000 cancer + 1,000 healthy)  & 400 & 0.7102 & 0.7125	&	0.7192 & 	0.7125 \\
\hline
\multirow{2}{*}{Authentic vs.~pseudo}
& \textbf{eccDNAMamba} & 20,000 (10,000 authentic + 10,000 pseudo) & 4,000 & \textbf{0.7401} & \textbf{0.7407}	& \textbf{0.7428} & \textbf{0.7407} \\
& DeepCircle (zero-shot)            & 20,000 (10,000 authentic + 10,000 pseudo) & 4,000 & 	0.6363	& 0.6532 & 0.6883 &	0.6532 \\
& DeepCircle (fine-tuned)            & 20,000 (10,000 authentic + 10,000 pseudo) & 4,000 & 0.6712	& 0.6742	& 0.6808	& 0.6742 \\
\hline
\end{tabular}
}
\caption{Performance comparison of eccDNAMamba and DNABERT-2 under different training settings.}
\label{tab:eccdna-results}
\end{table*}

\subsection{Classifying eccDNA of Diseased and Healthy Origins}
\label{sec:task1}

To evaluate the utility of eccDNAMamba foundation model, we fine-tuned it on a binary classification task distinguishing eccDNA sequences derived from cancerous tissues versus those from healthy controls. The biological rationale for this downstream task stems from mounting evidence that eccDNAs play distinct roles in oncogenesis—often carrying amplified oncogenes or regulatory elements that drive tumor growth—while eccDNAs in normal cells tend to represent background genomic rearrangements or byproducts of DNA repair. From a utility perspective, a model capable of discriminating these contexts could therefore serve as both a diagnostic tool and a means to prioritize functionally relevant eccDNAs for experimental follow-up.

We first assessed model performance sequences shorter than 10kbp, where most prior models have been evaluated Upon training on a balanced dataset comprising 10,000 cancer and 10,000 healthy eccDNA sequences, and evaluating on an independent balanced test set of 4,000 sequences, our fine-tuned eccDNAMamba model achieved an accuracy, macro F1, and recall of 0.8242, surpassing all baseline models on these metrics. While its precision, 0.8242, was slightly lower than that of Caduceus, 0.8248, the overall performance, as recorded in Table~\ref{tab:eccdna-results}, demonstrates strong generalization ability and state-of-the-art effectiveness in eccDNA classification.

To test the limits of generalization, we further evaluated all models on ultra-long eccDNA sequences (10kbp–200kbp), which pose greater computational and modeling challenges due to extended dependencies and sparse signal patterns. Despite the increased sequence complexity, eccDNAMamba retained its performance advantage.  As shown in Table~\ref{tab:eccdna-results}, fine-tuned eccDNAMamba achieves consistent state-of-the-art performance on ultra-long eccDNA sequences (10–200kbp) across all four evaluation metrics. It reaches an accuracy of 0.8175, macro F1 of 0.8147, recall of 0.8174, and precision of 0.8377, indicating both strong overall correctness and robust sequence-level discriminative capacity.

In contrast, DNABERT-2, which was truncated to the first 10kbp as recommended by the official model, suffered substantial degradation across all metrics. It achieved only 0.5702 macro F1, with accuracy, precision, and recall all hovering around 0.57.

HeynaDNA, while capable of consuming the full 200kbp inputs, produced moderately strong results (F1 = 0.7261, accuracy = 0.7350, precision = 0.7699, recall = 0.7350), but still underperformed compared to eccDNAMamba.

Caduceus was evaluated under two practical settings, based on its architecture, which processes input at base resolution, treating each nucleotide as a single token. 
In the first setting, we input a 131kbp base-level sequence, following the model’s reported maximum input length. 
In the second setting, we input 30,000 BPE tokens into our model, and simultaneously provided 30,000 bases (i.e., 30,000 base-level tokens) to Caduceus. 
This choice aligns the number of tokens between the two models. So both models receive an equal representational budget in terms of token count, facilitating a fair comparison.
These two Caduceus settings—one with 131,000 bases and the other with 30,000 bases (token-matched to our 30,000 BPE tokens)—yielded macro F1 scores of 0.7102 and 0.6922, respectively, with consistently lower accuracy, precision, and recall than eccDNAMamba.

These results highlight the importance of the ability to handle ultra-long sequence and modeling entire eccDNA sequences in their full ultralong context. Unlike baseline models that rely on truncation or base-equivalent approximations, eccDNAMamba processes complete sequences end-to-end, resulting in superior performance across all classification metrics. eccDNAMamba is the only model that remains robust across all evaluation metrics when handling ultra-long eccDNA sequences.

\subsection{eccDNAMamba Differentiates eccDNA from Genomic Pseudo-circular DNA}
\label{sec:task2}

A comprehensive understanding of the origins and sequence characteristics of extrachromosomal circular DNAs (eccDNAs) remains incomplete. In particular, it is unknown whether authentic eccDNAs exhibit distinguishable sequence features or merely resemble random circularized genomic fragments. To address this, we formulated a binary classification task comparing true eccDNAs against pseudo-circular DNA sequences—simulated fragments randomly extracted from the reference genome, carefully sampled to match the length distribution of authentic eccDNAs. These examples do not overlap with known eccDNAs or low-complexity (“N”) regions, and underwent identical preprocessing, including 64-token circular wrapping.

We fine-tuned eccDNAMamba on a balanced dataset of 10,000 true eccDNAs and 10,000 pseudo-circular fragments. Importantly, our model ingested full-length sequences without truncation, enabling it to leverage contextual signals up to 200kbp. On an independent test set of 4,000 sequences, eccDNAMamba achieved an accuracy of 0.7407, macro F1 of 0.7401, precision of 0.7428, and recall of 0.7407, demonstrating that authentic eccDNAs encode learnable, non-random sequence patterns—even under structural and length-matched controls.

We compared this performance to DeepCircle~\cite{10.1093/bib/bbad147}, a domain-specific SOTA model developed for eccDNA detection. DeepCircle employs a convolutional neural network(CNN) architecture and accepts only fixed-length 1000bp inputs, limiting its ability to model long-range dependencies. In the zero-shot setting, DeepCircle attained a macro F1 of 0.6363, and after task-specific fine-tuning, improved to 0.6712. In contrast, eccDNAMamba outperformed DeepCircle by over 6.9 percentage points in F1 and 6.6 points in accuracy (Table~\ref{tab:eccdna-results}. As reported in section~\ref{sec: model}, eccDNAMamba does not include a built-in classification head and is therefore not applicable to zero-shot evaluation in its pretrained form. As a pre-trained model, it requires task-specific fine-tuning to enable downstream tasks.
These results highlight two key insights: first, that authentic eccDNAs carry consistent sequence-level signals that distinguish them from simulated pseudo-circular fragments; and second, that long-context modeling and general-purpose sequence representations, as offered by eccDNAMamba, are critical for uncovering such patterns. The model’s strong performance against a domain-specific SOTA baseline underscores its potential as a versatile foundation for diverse eccDNA-related tasks.

\section{Analysis}

\begin{figure}[htbp]
    \centering
    \includegraphics[width=0.9\columnwidth]{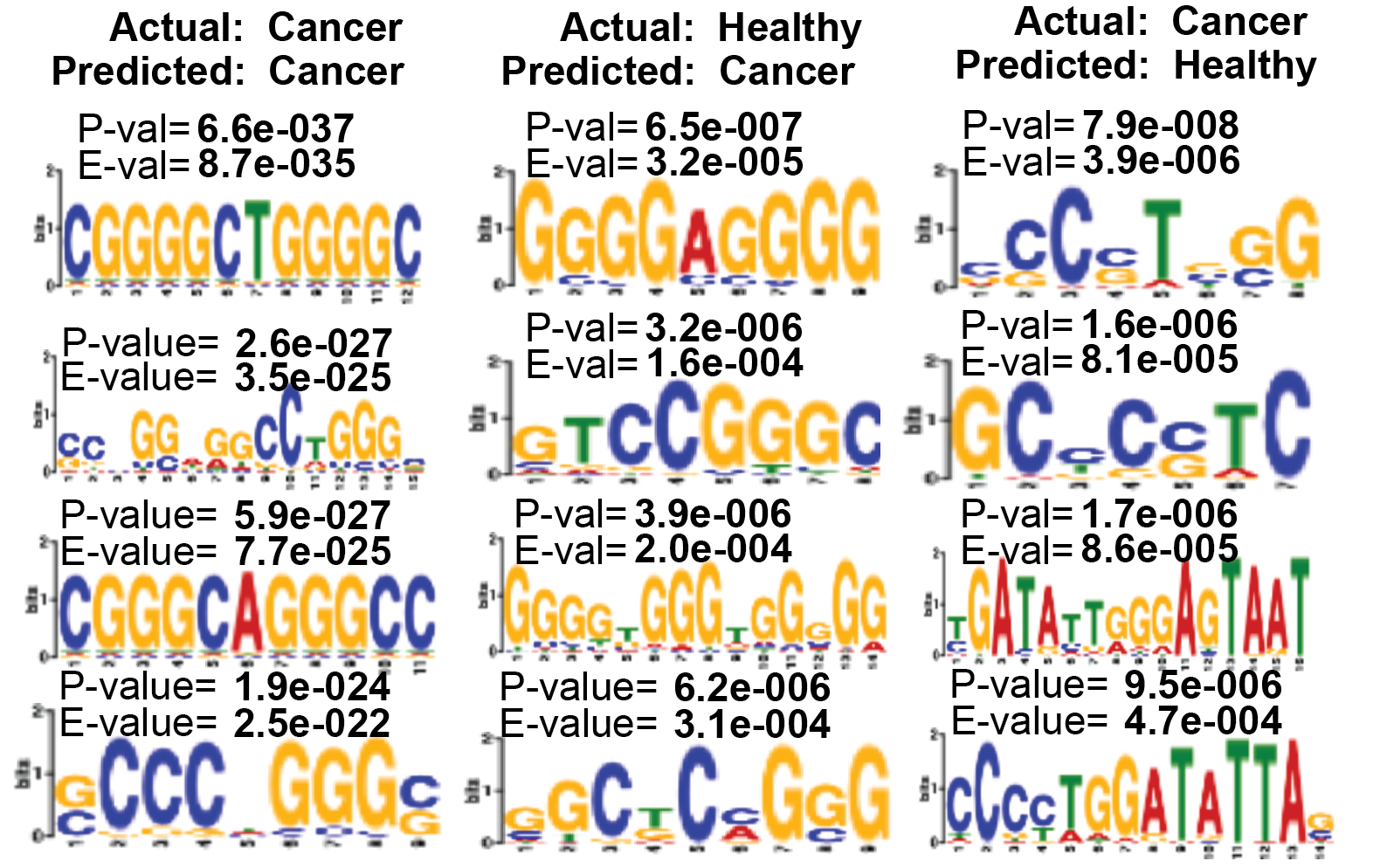}
    \caption{Top motifs in correctly predicted cancer eccDNA (TP), healthy eccDNA predicted to be cancer(FP), and cancer eccDNA predicted to be healthy(FN) sequences. 
    }
    \label{fig:example_motif}
\end{figure}

To better understand the sequence patterns that enable our model to discriminate cancer–derived eccDNA from healthy controls, we performed analysis based on section ~\ref{sec:task1}. We subjected the correctly predicted cancer (TP) and healthy (TN) eccDNA sequences to MEME-Suite’s \texttt{STREME} workflow ~\cite{dis1}. 
Across $28,100$ TP sequences using $29,292$ TN as background, the search converged on 128 statistically significant motifs \(\bigl(P \le 5 \times 10^{-2}\bigr)\).  
Many of the top motifs display sharply peaked position-weight matrices centered on CG-rich cores, a hallmark of DNA-binding zinc-finger (ZF) domains ~\cite{dis2}.
A parallel analysis of the 6,708 healthy eccDNA sequences incorrectly predicted to cancer (FP) yielded highly similar CG-rich motifs, whereas the 7,900 cancer eccDNA sequences incorrectly predicted to healthy (FN) produced distinct, AT-rich patterns.(Figure~\ref{fig:example_motif})  

This contrast suggests that our classifier relies on biologically relevant ZF-associated motifs; a minority of cancer eccDNAs lacking this signature therefore slip past its decision boundary.
To identify the proteins recognizing these dominant CG-rich motifs, we queried the 128-motif library with \texttt{Tomtom} against the CIS-BP\,2.0 Human database ~\cite{weirauch2014cisbp}. 
\texttt{Tomtom} returned 568 transcription-factor matches, 218 of which belong to the C2H2 ZF family~\cite{dis3}.  
We prioritized candidates by ranking them on combined motif occurrence and match significance (sum of \(-\log_{10}\)~E-values).  
This ranking is headed by \textit{ZNF24} and \textit{ZNF263}—genes implicated in cell-proliferation control, oncogene regulation, and multiple cancers—reinforcing a mechanistic link between ZF–DNA interactions and eccDNA dynamics in oncogenesis ~\cite{dis4,dis5}. 
Gene-set enrichment of the ranked list (GSEA-Preranked, MSigDB~v2024.1, C6 oncogenic signatures) ~\cite{dis} revealed one enriched group, \texttt{STK33\_SKM\_UP} \(\bigl(\text{norm.\ }p = 0.003,\; \text{FDR} = 0.110\bigr)\).

Taken together, these findings indicate that CG-rich C2H2 ZF motifs form the dominant sequence signature guiding our classifier’s cancer calls, whereas eccDNAs that evade detection (the AT-rich FN set) likely follow an alternative regulatory logic that future model iterations may need to capture.

\section{Discussion}
To enable scalable modeling and biological interpretation of ultra-long extrachromosomal circular DNA (eccDNA) sequences, we developed eccDNAMamba, a foundation model capable of capturing structural and regulatory patterns across sequence lengths up to 200k bp. To our knowledge, this is the first bidirectional state-space encoder tailored for circular DNA. The modular framework of eccDNAMamba allows extension to other downstream tasks such as eccDNA origin prediction and disease condition detection. Its ability to process full-length eccDNAs without truncation opens opportunities to study ultralong sequence organization and fine-grained motif syntax in circular DNA.
Compared to existing genome pretraining models, eccDNAMamba is pretrained on only 100 million tokens from unlabeled eccDNA sequences and fine-tuned on cancer-related classification tasks. Despite having significantly fewer pretraining tokens than Caduceus (35 billion) and DNABERT-2 (262 billion), eccDNAMamba consistently outperforms these foundation models across multiple classification benchmarks.  Importantly, the model shows no performance degradation with increasing sequence length, highlighting its effectiveness in eccDNA modeling and robust generalization in ultra-long eccDNA sequences. 

Our analysis reveals that the model's cancer-prediction decisions are guided by CG-rich sequence motifs resembling zinc finger (ZF) transcription factor binding sites. These motifs are enriched in both correctly predicted (TP) and misclassified (FP) cancer sequences, while false negatives (misclassified cancer samples) lack these CG-rich features and exhibit AT-rich signatures, indicating an alternative regulatory logic. Motif-to-factor mapping using Tomtom and CIS-BP further links the dominant motifs to C2H2 ZF proteins such as ZNF24 and ZNF263, known regulators of oncogenic pathways, reinforcing a mechanistic hypothesis specific to cancer eccDNA.

Nonetheless, the current design has limitations. The model’s reliance on CG-rich motifs may lead to blind spots, as indicated by the AT-rich false negatives. Future models may benefit from explicitly learning motif diversity or integrating external regulatory annotations. In future work, we aim to explore multimodal representations that combine sequence with regulatory profiles, further extend interpretability through attribution and counterfactual generation, and investigate the functional implications of model-derived motifs in experimental settings. We anticipate that eccDNAMamba will serve as a foundation for the broader understanding of eccDNA biology and as a generalizable framework for modeling long, circular genomic structures.

\section*{Acknowledgment}
We sincerely thank Dr. Ritambhara Singh and Jiaqi Zhang for their valuable feedback and insightful suggestions during the writing process that greatly improved this work. We also appreciate the anonymous reviewers for their constructive comments and recommendations during the review process.
\nocite{langley00}

\bibliography{example_paper}
\bibliographystyle{icml2025}

\newpage
\appendix
\onecolumn
\section{Supplementary Information}

\subsection{Interpreting Circular Augmentation}
\label{sec:ablation-1}

\begin{table}[ht]
\centering
\renewcommand{\arraystretch}{1.2}
\resizebox{\textwidth}{!}{
\begin{tabular}{llcccc}
\hline
\textbf{Architecture} & \textbf{Condition} & \textbf{Accuracy} & \textbf{F1 Score} & \textbf{Precision} & \textbf{Recall} \\
\hline
\multirow{2}{*}{CNN} 
& \textbf{With Circular Augmentation}    & \textbf{0.7985 ± 0.0025} & \textbf{0.7979 ± 0.0026} & \textbf{0.8023 ± 0.0033} & \textbf{0.7985 ± 0.0025} \\
& Without Circular Augmentation          & 0.7914 ± 0.0024 & 0.7906 ± 0.0022 & 0.7960 ± 0.0069 & 0.7914 ± 0.0024 \\
\hline
\multirow{2}{*}{MLP} 
& \textbf{With Circular Augmentation}    & \textbf{0.8002 ± 0.0015} & \textbf{0.7995 ± 0.0022} & \textbf{0.8041 ± 0.0027} & \textbf{0.8002 ± 0.0015} \\
& Without Circular Augmentation          & 0.7906 ± 0.0003 & 0.7900 ± 0.0010 & 0.7942 ± 0.0039 & 0.7906 ± 0.0003 \\
\hline
\multirow{2}{*}{Mamba} 
& \textbf{With Circular Augmentation}    & \textbf{0.8048 ± 0.0065} & \textbf{0.8046 ± 0.0066} & \textbf{0.8064 ± 0.0056} & \textbf{0.8048 ± 0.0065} \\
& Without Circular Augmentation          & 0.8011 ± 0.0021 & 0.8009 ± 0.0022 & 0.8024 ± 0.0016 & 0.8011 ± 0.0021 \\
\hline
\end{tabular}
}

\caption{Effect of Circular Augmentation on Model Performance for short eccDNA sequences ($<$200\,bp). Each model was trained for 3 epochs, and performance was evaluated on the test set using the checkpoint from the final epoch. Training and evaluation were repeated with 3 different random seeds, and the reported results represent the average across these runs.}
\label{tab:circular-augmentation}
\end{table}
\vspace{-5mm}
Due to computational resource limitations, we were unable to conduct an ablation study through repeated pretraining to systematically evaluate the effect of circular augmentation. As a practical alternative, we designed a preliminary experiment prior to pretraining to explore whether circular augmentation provides a meaningful benefit. Specifically, we evaluated its impact using Convolutional Neural Network (CNN), Multilayer Perceptron (MLP), and Mamba architectures on the classification task described in Section~\ref{sec:task2}.

Unlike the original task setup, we restricted the test set to eccDNA sequences shorter than 200 base pairs. All sequences were one-hot encoded, and circular augmentation was applied by appending the full-length eccDNA sequence to itself, rather than a fixed-length segment.

For the CNN model, we used a structure consisting of two convolutional layers (each with 16 filters and a kernel size of 3, followed by ReLU activation and L2 regularization), interleaved with dropout layers (rate 0.1), followed by a max pooling layer, a flattening layer, a dense layer with 100 units, and a final softmax classification layer. The model was optimized using the Adam optimizer with a learning rate of 1e-3.

The MLP model began with a flattening layer, followed by three fully connected layers with 512, 256, and 100 units respectively. Each dense layer was activated with ReLU and regularized with L2, and dropout was applied at rates of 0.3, 0.2, and 0.1 respectively. The final layer used softmax activation to output class probabilities. This model was trained using Adam with a learning rate of 5e-3.

The Mamba-based model consisted of a linear projection layer that mapped the input dimension to a 64-dimensional space, followed by a Mamba layer configured with a model dimension of 64, state size of 16, an expansion factor of 2, and an automatically selected DT rank. After feature extraction, adaptive average pooling was applied, and classification was performed using a two-layer fully connected network (a 100-unit ReLU layer with dropout, followed by an output layer).

Each model was trained for 3 epochs, and evaluation was performed on the test set using the model state from the third epoch. The training and evaluation process was repeated three times per model, and the final performance was reported as the average across these runs.

Through these experiments, we observed that circular augmentation consistently improved performance across all models as shown in Table ~\ref{tab:circular-augmentation}. As a result, we chose to incorporate circular augmentation into our pretraining pipeline to enhance downstream model effectiveness.
\newpage
\subsection{Interpreting Dataset}
\label{sec:data-1}

\begin{figure}[htb]
    \centering
    \includegraphics[width=1\textwidth]{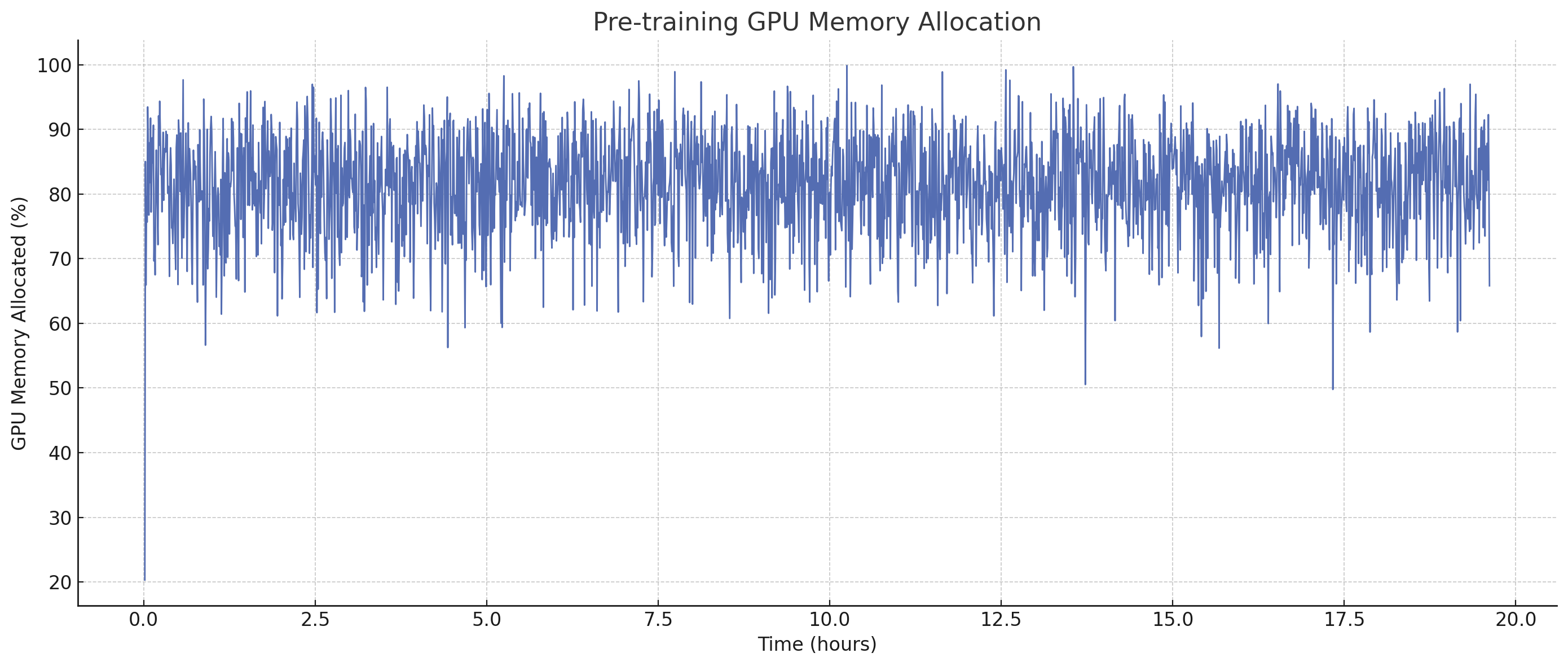}
    \caption{
        GPU memory utilization during pre-training of the eccDNAMamba model, measured as a percentage over time.
    }
    \label{fig:gpu_memory_allocation}
\end{figure}


Due to computational and memory constraints as shown in Figure ~\ref{fig:gpu_memory_allocation}, we did not adopt the strategy of appending the full-length eccDNA sequence to itself. Instead, we empirically chose to append approximately 25\% of the sequence to its end to ensure successful pre-training.

\begin{figure}[htb]
    \centering
    \subfigure[Token length distribution after BPE tokenization.]{
        \includegraphics[width=0.45\textwidth]{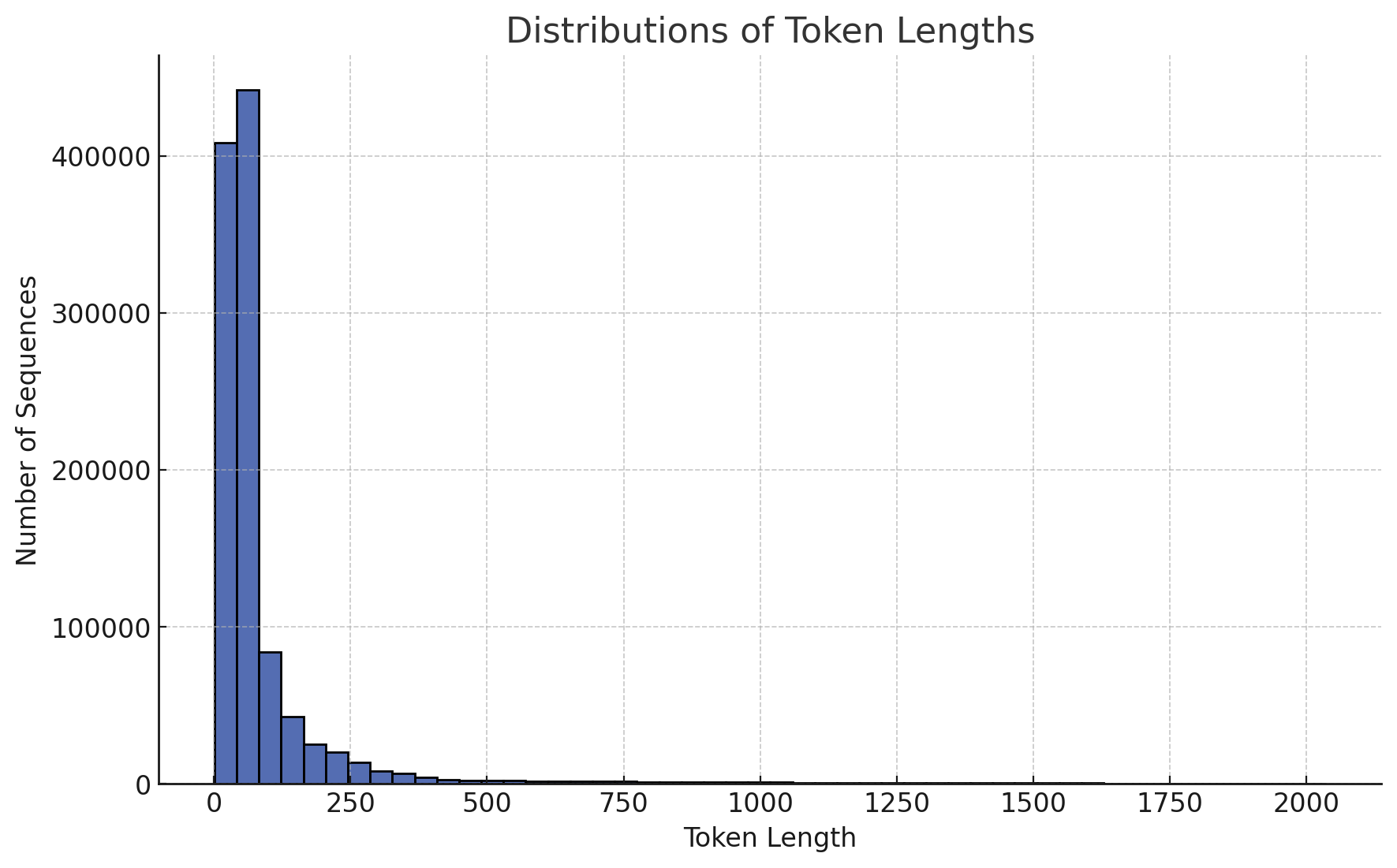}
        \label{fig:token_length_distribution}
    }
    \hspace{0.05\textwidth}
    \subfigure[Nucleotide vs. tokenized length.]{
        \includegraphics[width=0.45\textwidth]{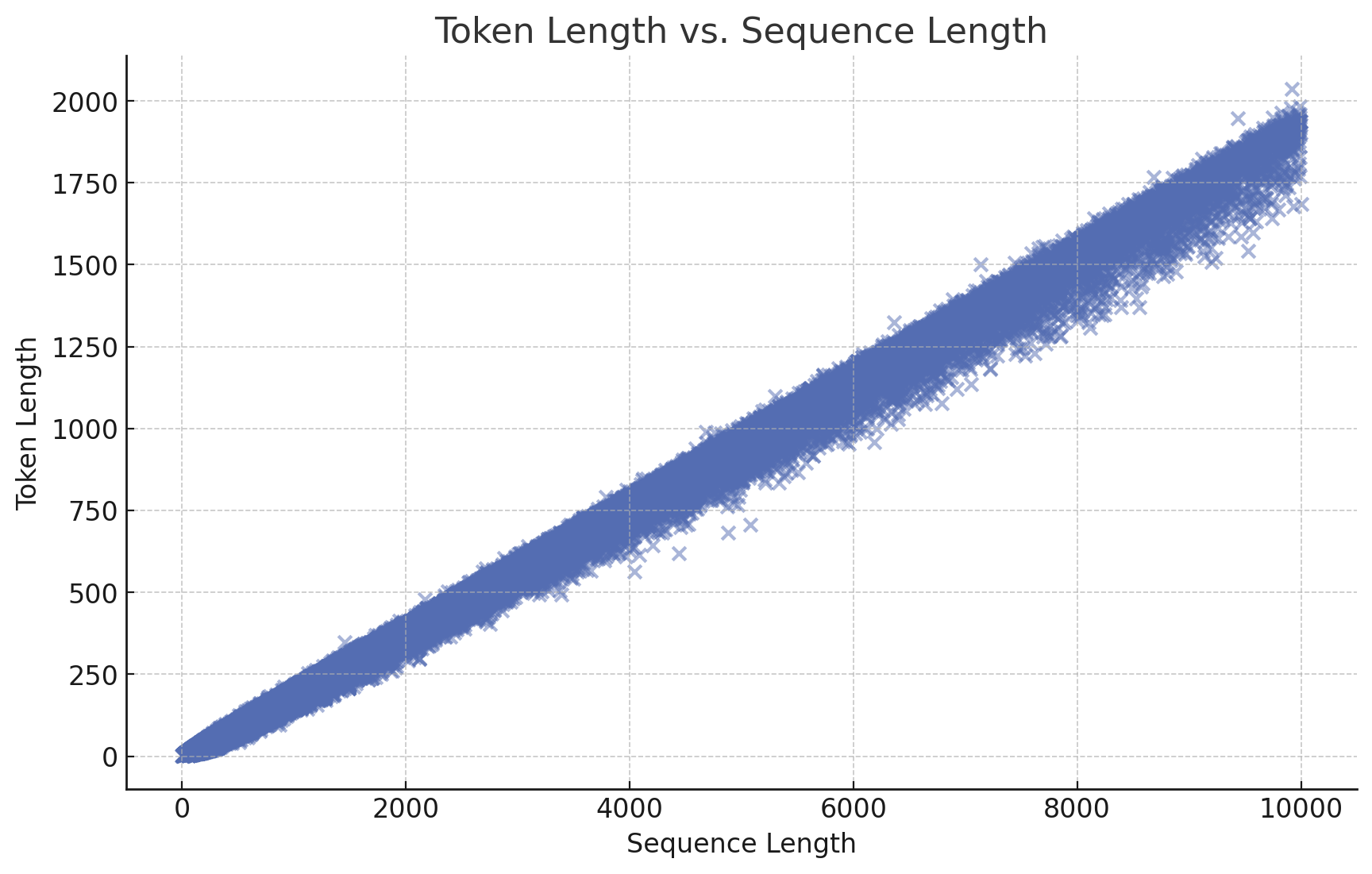}
        \label{fig:token_vs_sequence_length}
    }
    \caption{
        (a) Distribution of token lengths after BPE tokenization across the entire eccDNA dataset.
        Most sequences are tokenized into fewer than 256 tokens, with a long-tailed distribution extending to over 1000 tokens.
        (b) Relationship between nucleotide sequence length (in base pairs) and tokenized sequence length.
    }
\end{figure}

As shown in Figure~\ref{fig:token_length_distribution}, most eccDNA sequences in our dataset contain fewer than 256 tokens, so appending 25\% corresponds to 64 tokens.
\newpage
\subsection{Interpreting Byte Pair Encoding}
\label{sec:data-2}



We examined the compression efficiency of BPE by evaluating the relationship between raw nucleotide sequence length and the resulting tokenized length. As shown in Figure~\ref{fig:token_vs_sequence_length}, BPE substantially reduces token count across sequences of varying lengths, facilitating longer sequence modeling within the same context window. Specifically, after filtering out sequences longer than 10kbp, we obtained a final corpus of 1,087,886 eccDNA sequences totaling 524 million base pairs. Applying BPE tokenization to this dataset yielded approximately 101.5 million tokens, corresponding to an average of 5.16 base pairs per token.


You can have as much text here as you want. The main body must be at most $8$ pages long.
For the final version, one more page can be added.
If you want, you can use an appendix like this one.  

The $\mathtt{\backslash onecolumn}$ command above can be kept in place if you prefer a one-column appendix, or can be removed if you prefer a two-column appendix.  Apart from this possible change, the style (font size, spacing, margins, page numbering, etc.) should be kept the same as the main body.

\end{document}